\documentclass[12pt]{article}
\usepackage{amsmath}
\usepackage{amssymb}

\input{tcilatex}
\begin{document}

\begin{center}
\textbf{TOWARDS A UNIFIED LAGRANGIAN FORMALISM}

\smallskip \ 

\textbf{FOR COSMOLOGY AND BLACK HOLES}

\smallskip \ 

G. Avila$^{\dag }$\footnote[1]{%
a214290123@alumnos.unison.mx}, J. A. Nieto$^{\ddag }$\footnote[2]{%
niet@uas.edu.mx; janieto1@asu.edu}

\bigskip \ 

$^{\dag }$\textit{Departamento de Investigaci\'{o}n en F\'{\i}sica de la
Universidad}

\textit{de Sonora, C.P. 83000, Hermosillo, Sonora, M\'{e}xico}

\bigskip \ 

$^{\ddag }$\textit{Facultad de Ciencias F\'{\i}sico-Matem\'{a}ticas de la
Universidad Aut\'{o}noma}

\textit{de Sinaloa, C.P. 80010, Culiac\'{a}n Sinaloa, M\'{e}xico}

\smallskip

\bigskip \ 

\textbf{Abstract}
\end{center}

Using a Lagrangian formalism we establish a relationship between the $%
(n+D+d)-$dimensional cosmology, black-holes and the Polyakov action for
strings. Specifically, we identify these physical scenarios as part of a
2-dimensional metric, which arises from a Lagrangian function with
constraints, derived from the Einstein-Hilbert action. In particular, we
show that the Friedmann-Robertson-Walker cosmological model and the
Schwarzschild solution are both consequence of this Lagrangian.

\bigskip \ 

\bigskip \ 

\bigskip \ 

\ 

Key words; Cosmology, black-holes, string theory.

Pacs No.: 98.80.Dr, 12.10.Gq, 04.50.+h.

May 2018

\newpage

\section{\textbf{Introduction}}

\qquad The purpose of this work is to find a relationship between the
Friedmann-Robertson-Walker (FRW) metric, the Schwarzschild black hole
solution and strings. By considering that the first is a dynamical metric
and the second is static, as a preliminary step, we will get a relationship
between a black hole-type metric such as the de Sitter metric and the
FRW-metric.

As it has been recently shown, the FRW-cosmological model and the black-hole
formalism can be derived from a first order Lagrangian associated with a
constrained system (see Refs. [1]-[3]). In fact, by considering a particular
ansatz for the metric such a Lagrangian is obtained from the
Einstein-Hilbert action. In this work we shall show that considering a
similar procedure it is possible to unify cosmology and black-holes
concepts, which can be obtained as a limit case of this more general
constrained Lagrangian formalism. It turns out that our proposed Lagrangian
formalism can be associated with the Polyakov action in (1+1)-dimensions.

Our approach may be physically interesting for a number of reasons. First,
one may use the complete mathematical tools of Lagrangians for constrained
systems in order to study a number of symmetries underlying the cosmology
and black-hole solutions. Second, a unified treatment of cosmology and
black-holes may be of particular interest in the context of string theory or
M-theory (See Refs. [4]and [5]). One reason for this is because M-theory
predicts, among other things, that our universe can be described by a
brane-world [6]. So, it appears attractive to search for a
brane-world/black-hole correspondence (from the point of view of M-theory).
In fact, this is the main motivation to extend our approach in order to find
a more general formulation that links not only cosmology and black holes but
also string theory via the Polyakov action.

It is important to mention that the idea of combining the cosmology and
black-hole concepts has a long history (see Ref. [7] and references
therein). Let us just mention that perhaps one of the early ideas is the
so-called Swiss cheese universe which is constructed by cutting out spheres
from cosmology universe and collapsing the matter down into black holes [8].
At present, the subject is known as cosmological black holes (or
Schwarzschild cosmology, or black-hole cosmological model (see Ref. [7] and
references therein)). In any case, the main idea of these approaches is to
combine the Hubble radius $r_{H}$ of the observable universe with its
Schwarzschild radius $r_{S}$ of a black-hole [9]-[10]. The main difference
between our formalism and those previous works is that in our approach both
cosmology and black-hole are unified through a dynamic two dimensional
metric without requiring any combination between the radii $r_{H}$ and $%
r_{S} $.

The structure of this article is as follows. In section 2, we develop an
appropriate coordinate transformation to link a Schwarzschild-de Sitter type
metric with the FRW-metric. In section 3 and 4, we show that the
FRW-cosmological model and the Schwarzschild black-hole, respectively, are
particular cases of a more general first order Lagrangian associated with a
constrained system. In section 5, we derive the Polyakov action in 2
dimensions as another particular case for the same Lagrangian function.
Meanwhile, in section 6, we show the procedure can be generalized to any
dimension. Finally, in section 7, we make some final remarks.

\bigskip

\section{\textbf{Static and Non-static forms of the de Sitter space}}

\qquad Interpretational problems of the correspondence between static and
non-static de Sitter metrics via coordinate transformations have been
subject of interest [11]. In this section, the principal idea is to find a
link between black holes and cosmology at a level of coordinate
transformations. For this purpose, in order to clarify the formalism, we
will generalize our analysis beyond the de Sitter space.

Assuming a flat space for slices at comoving time constant, we may use the
usual FRW-cosmology line element

\begin{equation}
dS_{(1)}^{2}=-dT^{2}+a^{2}(T)(d\rho ^{2}+\rho ^{2}d\Omega ^{2}),  \tag{1}
\end{equation}%
where $T$ and $\rho $ are time and radial type coordinates and $d\Omega
^{2}=d\theta ^{2}+\sin ^{2}\theta $ $d\phi ^{2}$ is the solid angle line
element.

On the other hand, motivated by the vacuum solution for a spherical
symmetric source, one can propose a static line element of the form

\begin{equation}
dS_{(2)}^{2}=-f(r)dt^{2}+\frac{dr^{2}}{f(r)}+r^{2}d\Omega ^{2},  \tag{2}
\end{equation}%
with $t$ and $r$ being the usual time and radial coordinates. The question
arises: what kind of coordinate transformation can relate these two line
elements? In what follows we shall find a coordinate transformation which
connects the line elements (1) and (2).

By setting $dS_{(1)}^{2}=g_{\mu \nu }dx^{\mu }dx^{\nu }$ and $%
dS_{(2)}^{2}=\gamma _{ab}dy^{a}dy^{b}$ one finds the metric transformation

\begin{equation}
g_{\mu \nu }=\frac{\partial y^{a}}{\partial x^{\mu }}\frac{\partial y^{b}}{%
\partial x^{\nu }}\gamma _{ab}.  \tag{3}
\end{equation}%
Note that, due to spherical symmetry, one can make the identification%
\begin{equation}
r=a\rho ,  \tag{4}
\end{equation}%
which means that $r=r(T,\rho )$.

Now, one finds that relations (1) through (4) lead to the equations

\begin{equation}
-1=-\left( \frac{\partial t}{\partial T}\right) ^{2}f+\frac{\rho ^{2}\dot{a}%
^{2}}{f},  \tag{5}
\end{equation}

\begin{equation}
a^{2}=-\left( \frac{\partial t}{\partial \rho }\right) ^{2}f+\frac{a^{2}}{f}
\tag{6}
\end{equation}%
and

\begin{equation}
0=-\frac{\partial t}{\partial T}\frac{\partial t}{\partial \rho }f+\frac{%
\rho a\dot{a}}{f},  \tag{7}
\end{equation}%
where $\dot{a}=\frac{da}{dT}$. Using these equations, our goal is to find $%
t=t(T,\rho )$. The set of equations (5) through (7) can be rewritten as

\begin{equation}
\left( \frac{\partial t}{\partial T}\right) ^{2}=\frac{1}{f^{2}}(f+\rho ^{2}%
\dot{a}^{2}),  \tag{8}
\end{equation}

\begin{equation}
\left( \frac{\partial t}{\partial \rho }\right) ^{2}=\frac{a^{2}}{f^{2}}(1-f)
\tag{9}
\end{equation}%
and

\begin{equation}
f^{4}\left( \frac{\partial t}{\partial T}\right) ^{2}\left( \frac{\partial t%
}{\partial \rho }\right) ^{2}=\rho ^{2}a^{2}\dot{a}^{2}.  \tag{10}
\end{equation}%
Then, substituting (8) and (9) into (10) one obtains

\begin{equation}
f=1-\rho ^{2}\dot{a}^{2},  \tag{11}
\end{equation}%
where now we get $f=f(T,\rho )$.

However substituing (11) into (8) one obtains%
\begin{equation}
\frac{\partial t}{\partial T}=\frac{1}{f},  \tag{12}
\end{equation}%
while using equation (11) and (9) one finds that

\begin{equation}
\frac{\partial t}{\partial \rho }=\frac{\rho a\dot{a}}{f}.  \tag{13}
\end{equation}

By integrating (13) one gets

\begin{equation}
t=-\frac{a}{2\dot{a}}\ln f+g(T),  \tag{14}
\end{equation}%
where we have considered $\frac{\partial f}{\partial \rho }=-2\rho \dot{a}%
^{2}$. Here, $g(T)$ is an arbitrary function of $T$.

Thus, from (14) it is not difficult to see that

\begin{equation}
\frac{\partial t}{\partial T}=\frac{\rho ^{2}a\ddot{a}}{f}-\frac{1}{2}\left(
1-\frac{a}{\dot{a}}\frac{\ddot{a}}{\dot{a}}\right) \ln f+\dot{g},  \tag{15}
\end{equation}%
where we have used (11) to find that $\frac{\partial f}{\partial T}=-2\rho
^{2}\dot{a}\ddot{a}$.

Therefore, equating (12) and (15) yields

\begin{equation}
-\frac{f}{2}\left( 1-\frac{a}{\dot{a}}\frac{\ddot{a}}{\dot{a}}\right) \ln
f+\rho ^{2}a\ddot{a}+\dot{g}f=1.  \tag{16}
\end{equation}

On the limit for a flat spacetime on (2), that is when $r\rightarrow \infty $%
, one must have $f(r)\rightarrow 1$ and we can see from (11) that this
implies $\rho ^{2}\dot{a}^{2}\rightarrow 0$. But acording to (4) one must
have $r\rightarrow 0$ and thus this means that one can set $\dot{g}%
(T)\rightarrow 1$. Therefore up to an integration constant one can set $%
g(T)=T$. Considering these results one finds that (16) can be rewriten as

\begin{equation}
\left( 1-\frac{a}{\dot{a}}\frac{\ddot{a}}{\dot{a}}\right) \left( \frac{f}{2}%
\ln f+1-f\right) =0.  \tag{17}
\end{equation}

Assuming that one has $\frac{f}{2}lnf+1-f=0$ we notice that this is
satisfied only if one sets $f=1$, which as we have mentioned earlier it is
not possible in general. Hence one must set the first factor in (17) equal
to zero, and then 
\begin{equation}
\frac{\ddot{a}}{\dot{a}}=\frac{\dot{a}}{a},  \tag{18}
\end{equation}%
which implies that $\frac{d}{dt}\ln (\frac{\dot{a}}{a})=0$, and therefore 
\begin{equation}
a(T)=e^{T/r_{0}},  \tag{19}
\end{equation}%
where $r_{0}$ is an integration constant. From this result on finds that
(14) becomes

\begin{equation}
t(T,\rho )=-\frac{r_{0}}{2}\ln (1-\frac{e^{2T/r_{0}}\rho ^{2}}{r_{0}^{2}})+T,
\tag{20}
\end{equation}%
and we see that (4) can be rewritten as%
\begin{equation}
r(T,\rho )=e^{T/r_{0}}\rho .  \tag{21}
\end{equation}%
\bigskip

Equation (19) also implies that (1) can be written as%
\begin{equation}
dS_{(1)}^{2}=-dT^{2}+e^{2T/r_{0}}(d\rho ^{2}+\rho ^{2}d\Omega ^{2}), 
\tag{22}
\end{equation}%
and it is easy to see that (11) can be written as follows 
\begin{equation}
f=1-\frac{r^{2}}{r_{0}^{2}},  \tag{23}
\end{equation}%
where now we have $f=f(r)$.

This means that we can rewrite (2) as follows%
\begin{equation}
dS_{(2)}^{2}=-\left( 1-\frac{r^{2}}{r_{0}^{2}}\right) dt^{2}+\frac{dr^{2}}{%
\left( 1-\frac{r^{2}}{r_{0}^{2}}\right) }+r^{2}d\Omega ^{2}.  \tag{24}
\end{equation}%
\bigskip \ 

Equations (20) and (21) are the transformations from $(T,\rho )$ system to $%
(t,r)$, that allows to go from (24) to (22). The inverse transformations
comes from (20) and (21) directly, which imply

\begin{equation}
T(t,r)=\frac{r_{0}}{2}\ln \left( 1-\frac{r^{2}}{r_{0}^{2}}\right) +t, 
\tag{25}
\end{equation}%
while $\rho $ is obtained from (21) and (25)

\begin{equation}
\rho (t,r)=\frac{r}{\sqrt{1-\frac{r^{2}}{r_{0}^{2}}}}e^{-t/r_{0}},  \tag{26}
\end{equation}%
which allows to go from (22) to (24).

It is remarkable to mention that based on our analysis, if one wants to find
a coordinate transformation between (1) and (2), the only possible solution
is the de Sitter space, and our equations (25) and (26) are reduced to the
already known case [12]. That is, this demonstrates the uniqueness of the
solution when one imposes, via coordinate transformations, equivalence
between the metrics (1) and (2).

\bigskip

\section{\textbf{Cosmology as a projection of a general }$(n+D+d)-$%
dimensional \textbf{Einstein-Hilbert} \textbf{action.}}

\qquad We now consider the Einstein-Hilbert action in $(n+D+d)-$dimensions,%
\begin{equation}
S=\frac{1}{V_{D+d}}\int d^{n+D+d}x\sqrt{-g}(R-2\Lambda ),  \tag{27}
\end{equation}%
and we assume the anzats

\begin{equation}
ds^{2}=\bar{g}_{AB}(x^{C})dx^{A}dx^{B}+a^{2}(x^{C})d^{D}\Omega
+b^{2}(x^{C})d^{d}\Sigma ,  \tag{28}
\end{equation}%
where the indices $A,B$ run from $1$ to $n$. Here, the expressions $%
d^{D}\Omega \equiv \tilde{g}_{ij}(x^{k})dx^{i}dx^{j}$ and $d^{d}\Sigma
\equiv \hat{g}_{ab}(x^{d})dx^{a}dx^{b}$ correspond to a $D$-dimensional and $%
d$-dimensional maximally symmetric homogeneous spaces, with constant
curvature $k_{1}=0,\pm 1$ and $k_{2}=0,\pm 1$, respectively.

It can be shown that using the line element (28), the action (27) is reduced
to

\begin{equation}
\begin{array}{c}
S=\int d^{n}x\sqrt{-\bar{g}}a^{D}b^{d}\{-2Da^{-1}\mathcal{D}_{A}\partial
^{A}a-D(D-1)\bar{g}^{AB}a^{-2}\partial _{A}a\partial _{B}a \\ 
\\ 
-2db^{-1}\mathcal{D}_{A}\partial ^{A}b-d(d-1)\bar{g}^{AB}b^{-2}\partial
_{A}b\partial _{B}b \\ 
\\ 
-2Dd(b^{-1}a^{-1})\bar{g}^{AB}\partial _{A}a\partial _{B}b+\bar{R}+a^{-2}%
\tilde{R}+b^{-2}\hat{R}-2\Lambda \},%
\end{array}
\tag{29}
\end{equation}%
where $\bar{g}=\det (\bar{g}_{AB})$. Notice that we are choosing $\bar{g}%
_{AB}$ to be a Lorentzian metric.

The action (29) can be rewritten as

\begin{equation}
\begin{array}{c}
S=-\int d^{n}x\sqrt{-\bar{g}}\{ \mathcal{D}_{A}(2Da^{D-1}\partial
^{A}ab^{d}+2db^{d-1}\partial ^{A}ba^{D}) \\ 
\\ 
-D(D-1)\bar{g}^{AB}a^{D-2}b^{d}\partial _{A}a\partial _{B}a-d(d-1)\bar{g}%
^{AB}a^{D}b^{d-2}\partial _{A}b\partial _{B}b \\ 
\\ 
-2Dd(a^{D-1}b^{d-1})\bar{g}^{AB}\partial _{A}a\partial _{B}b-a^{D}b^{d}(\bar{%
R}+a^{-2}\tilde{R}+b^{-2}\hat{R}-2\Lambda )\},%
\end{array}
\tag{30}
\end{equation}%
where $\mathcal{D}_{A}$ is a covariant derivative associated with $\bar{g}%
_{AB}$. Dropping the total derivative in (30) one obtains,

\begin{equation}
\begin{array}{c}
S=\int d^{n}x\sqrt{-\bar{g}}a^{D}b^{d}\{D(D-1)\bar{g}^{AB}a^{-2}\partial
_{A}a\partial _{B}a+d(d-1)\bar{g}^{AB}b^{-2}\partial _{A}b\partial _{B}b \\ 
\\ 
+2Dd(a^{-1}b^{-1})\bar{g}^{AB}\partial _{A}a\partial _{B}b+(\bar{R}+a^{-2}%
\tilde{R}+b^{-2}\hat{R}-2\Lambda )\}.%
\end{array}
\tag{31}
\end{equation}%
Here, $\bar{R},\tilde{R}$ and $\hat{R}$ are the curvature scalars associated
with $\bar{g}_{AB}(x^{C}),$ $\tilde{g}_{ij}(x^{k})$ and $\hat{g}_{ab}(x^{d})$%
, respectively (see Ref. [1] and [2] for details).

It is worth mentioning that provided $n\neq 2$ and $\bar{R}+a^{-2}\tilde{R}%
+b^{-2}\hat{R}-2\Lambda =0$, one finds that the action (31) is invariant
under the duality transformation,

\begin{equation}
\begin{array}{ccc}
a & \rightarrow & \frac{1}{a}, \\ 
&  &  \\ 
b & \rightarrow & \frac{1}{b}, \\ 
&  &  \\ 
g_{AB} & \rightarrow & a^{\frac{4D}{n-2}}b^{\frac{4d}{n-2}}\bar{g}_{AB},%
\end{array}
\tag{32}
\end{equation}%
(see also Refs. [1], [13] and [14]). Therefore, from point of view of the
duality transformation (32), the model (31) with $2$-dimensional metric $%
\bar{g}_{AB}(x^{C})$ is an exceptional case. In some sense the duality
symmetry is playing here the analogue role as the Weyl invariance in $p$%
-brane physics (see Ref. [4] and references therein), which implies that the 
$1$-brane (string) theory is an exceptional case. For $n=2$ the action (31)
becomes

\begin{equation}
\begin{array}{c}
S=\int d^{2}x\sqrt{-\bar{g}}a^{D}b^{d}\{D(D-1)\bar{g}^{AB}a^{-2}\partial
_{A}a\partial _{B}a+d(d-1)\bar{g}^{AB}b^{-2}\partial _{A}b\partial _{B}b \\ 
\\ 
+2Dd(a^{-1}b^{-1})\bar{g}^{AB}\partial _{A}a\partial _{B}b+(\bar{R}+a^{-2}%
\tilde{R}+b^{-2}\hat{R}-2\Lambda )\}.%
\end{array}
\tag{33}
\end{equation}

Considering a particular case of (28) expressed by the line element

\begin{equation}
ds^{2}=-N^{2}(t)dt^{2}+a^{2}(t)d^{D}\Omega +b^{2}(t)d^{d}\Sigma ,  \tag{34}
\end{equation}%
the Einstein-Hilbert action (33) would be simplified to

\begin{equation}
\begin{array}{c}
S=\int dt\{N^{-1}a^{D}b^{d}[D(D-1)a^{-2}\dot{a}^{2}+d(d-1)b^{-2}\dot{b}%
^{2}+2dDa^{-1}\dot{a}b^{-1}\dot{b}] \\ 
\\ 
-D(D-1)k_{1}Na^{D-2}b^{d}-d(d-1)k_{2}Na^{D}b^{d-2}+2\Lambda Na^{D}b^{d}\},%
\end{array}
\tag{35}
\end{equation}%
where it is easy to show that the field equations for a cosmology model in $%
(1+D+d)-$dimensions follows from (34) and (35) [1].

The problem with (34) is that it can be obtained from (28) by taking $n=2$
and choosing $\bar{g}_{11}(x^{C})=-N^{2}$, $\bar{g}_{12}(x^{C})=\bar{g}%
_{21}(x^{C})=0$ and $\bar{g}_{22}(x^{C})=0$. This means that the $2$%
-dimensional metric $\bar{g}_{AB}(x^{C})$ in (33) is singular, in the sense
that $det[\bar{g}_{AB}(x^{C})]=0$. So, the next question arises; starting
from (28) and assuming $det[\bar{g}_{AB}(x^{C})]\neq 0$, how can the line
element (34) be obtained? The answer to this question may be solved by
performing a different type of projection. Let us rewrite the ansatz (28) in
the form

\begin{equation}
ds^{2}=\bar{g}_{AB}(x^{1},x^{2})dx^{A}dx^{B}+a^{2}(x^{1},x^{2})d^{D}\Omega
+b^{2}(x^{1},x^{2})d^{d}\Sigma ,  \tag{36}
\end{equation}%
and assuming $\bar{g}_{12}(x^{1},x^{2})=\bar{g}_{21}(x^{1},x^{2})=0$, one
obtains

\begin{equation}
ds^{2}=\bar{g}_{11}(x^{1},x^{2})dx^{1}dx^{1}+\bar{g}%
_{22}(x^{1},x^{2})dx^{2}dx^{2}+a^{2}(x^{1},x^{2})d^{D}\Omega
+b^{2}(x^{1},x^{2})d^{d}\Sigma .  \tag{37}
\end{equation}%
Now, by performing the projections

\begin{equation}
\begin{array}{c}
\bar{g}_{11}(x^{1},x^{2})\longrightarrow \bar{g}_{11}(x^{1})=-N^{2}, \\ 
\\ 
\bar{g}_{22}(x^{1},x^{2})\longrightarrow \bar{g}_{22}(x^{1})=a^{2}(x^{1}),
\\ 
\\ 
a^{2}(x^{1},x^{2})\longrightarrow a^{2}(x^{1}), \\ 
\\ 
b^{2}(x^{1},x^{2})\longrightarrow b^{2}(x^{1}),%
\end{array}
\tag{38}
\end{equation}%
the line element (37) becomes

\begin{equation}
ds^{2}=-N^{2}(x^{1})dx^{1}dx^{1}+a^{2}(x^{1})d^{1+D}\Omega
+b^{2}(x^{1})d^{d}\Sigma .  \tag{39}
\end{equation}%
Here, we have defined $d^{1+D}\Omega =dx^{2}dx^{2}+d^{D}\Omega $. Therefore,
one has discovered that (39) has exactly the same form as (34) and as a
consequence the action (35) follows, provided one makes the metric extension

\begin{equation}
\tilde{g}_{ij}\rightarrow \left( 
\begin{array}{cc}
1 & 0 \\ 
0 & \tilde{g}_{ij}%
\end{array}%
\right) .  \tag{40}
\end{equation}%
Moreover one can verify that for $n=2$ and the projection (38) applied to
(33) then the action (35) can be obtained.

\bigskip

\section{\textbf{Black Holes as a projection of a general }$(n+D+d)-$%
dimensional \textbf{Einstein-Hilbert} \textbf{action.}}

\qquad For the case of black-holes we now consider the following reductions

\begin{equation}
\begin{array}{c}
\bar{g}_{11}(x^{1},x^{2})\longrightarrow \bar{g}_{11}(x^{2})=-e^{f(r)}, \\ 
\\ 
\bar{g}_{22}(x^{1},x^{2})\longrightarrow \bar{g}_{22}(x^{2})=e^{h(r)}, \\ 
\\ 
a^{2}(x^{1},x^{2})\longrightarrow a^{2}(x^{2})\longrightarrow \varphi
^{2}(r), \\ 
\\ 
b^{2}(x^{1},x^{2})\longrightarrow b^{2}(r),%
\end{array}
\tag{41}
\end{equation}%
\bigskip where $x^{2}=r$, and one finds that in this case the line element
(37) becomes

\begin{equation}
ds^{2}=-e^{f(r)}(dx^{1})^{2}+e^{h(r)}dr^{2}+\varphi ^{2}(r)d^{D}\Omega
+b^{2}(r)d^{d}\Sigma ,  \tag{42}
\end{equation}%
which is the well known Schwarzschild metric in higher dimensions with
scalar curvature $R$ being%
\begin{equation}
\begin{array}{c}
R=e^{-h}\{-\ddot{f}-\frac{\dot{f}^{2}}{2}+\frac{\dot{f}\dot{h}}{2}+D(\dot{h}-%
\dot{f})\frac{\dot{\varphi}}{\varphi }-2D\frac{\ddot{\varphi}}{\varphi } \\ 
\\ 
-D(D-1)\frac{\dot{\varphi}^{2}}{\varphi ^{2}}+d(\dot{h}-\dot{f})\frac{\dot{b}%
}{b}-2d\frac{\ddot{b}}{b}-d(d-1)\frac{\dot{b}^{2}}{b^{2}}-2Dd\frac{\dot{%
\varphi}}{\varphi }\frac{\dot{b}}{b}\} \\ 
\\ 
+k_{1}D(D-1)\varphi ^{-2}+k_{2}d(d-1)b^{-2},%
\end{array}
\tag{43}
\end{equation}%
where the dot in any quantity $\dot{A}$ means derivative with respect $r$.
On the other hand, we have

\begin{equation}
\sqrt{-g}=e^{\frac{f+h}{2}}\varphi ^{D}b^{d}\sqrt{\tilde{g}}\sqrt{\hat{g}}, 
\tag{44}
\end{equation}%
where $g$, $\tilde{g}$ and $\hat{g}$ denote the determinant of $g_{\mu \nu }$%
, $\tilde{g}_{ij}$ and $\hat{g}_{ab}$, respectively. Consequently, one can
show that the higher dimensional Einstein-Hilbert action

\begin{equation}
S=\frac{1}{V_{D+d}}\int_{M^{(1+1)+D+d}}\sqrt{-g}R,  \tag{45}
\end{equation}%
can be simplified to the form%
\begin{equation}
\begin{array}{c}
S=\int dr\{(e^{\frac{f-h}{2}}\varphi ^{D}b^{d})[-\ddot{f}-\frac{\dot{f}^{2}}{%
2}+\frac{\dot{f}\dot{h}}{2}+D(\dot{h}-\dot{f})\frac{\dot{\varphi}}{\varphi }%
-2D\frac{\ddot{\varphi}}{\varphi } \\ 
\\ 
-D(D-1)\frac{\dot{\varphi}^{2}}{\varphi ^{2}}+d(\dot{h}-\dot{f})\frac{\dot{b}%
}{b}-2d\frac{\ddot{b}}{b}-d(d-1)\frac{\dot{b}^{2}}{b^{2}}-2Dd\frac{\dot{%
\varphi}}{\varphi }\frac{\dot{b}}{b}] \\ 
\\ 
+k_{1}D(D-1)e^{\frac{f+h}{2}}\varphi ^{D-2}b^{d}+k_{2}d(d-1)e^{\frac{f+h}{2}%
}\varphi ^{D}b^{d-2}\}.%
\end{array}
\tag{46}
\end{equation}

Furthermore, one can prove that (46) can be rewritten as

\begin{equation}
\begin{array}{c}
S=\int dr\{-2\frac{d}{dr}(e^{\frac{-h}{2}}\frac{d}{dr}(\varphi ^{D}b^{d}e^{%
\frac{f}{2}}))+(e^{\frac{f-h}{2}}\varphi ^{D}b^{d})[D(D-1)\frac{\dot{\varphi}%
^{2}}{\varphi ^{2}}+d(d-1)\frac{\dot{b}^{2}}{b^{2}} \\ 
\\ 
+\frac{D\dot{\varphi}\dot{f}}{\varphi }+\frac{d\dot{b}\dot{f}}{b}+2Dd\frac{%
\dot{b}}{b}\frac{\dot{\varphi}}{\varphi }\}] \\ 
\\ 
+k_{1}D(D-1)e^{\frac{f+h}{2}}\varphi ^{D-2}b^{d}+k_{2}d(d-1)e^{\frac{f+h}{2}%
}\varphi ^{D}b^{d-2}\}.%
\end{array}
\tag{47}
\end{equation}

Dropping the total derivative, the action (47) is reduced to%
\begin{equation}
\begin{array}{c}
S=\int dr\{(\Omega ^{-1}\mathcal{F}\varphi ^{D}b^{d})[D(D-1)\frac{\dot{%
\varphi}^{2}}{\varphi ^{2}}+d(d-1)\frac{\dot{b}^{2}}{b^{2}}+2D\frac{\mathcal{%
\dot{F}}}{\mathcal{F}}\frac{\dot{\varphi}}{\varphi }+2d\frac{\mathcal{\dot{F}%
}}{\mathcal{F}}\frac{\dot{b}}{b}+2Dd\frac{\dot{b}}{b}\frac{\dot{\varphi}}{%
\varphi }] \\ 
\\ 
+\Omega \mathcal{F}[k_{1}D(D-1)\varphi ^{D-2}b^{d}+k_{2}d(d-1)\varphi
^{D}b^{d-2}]\},%
\end{array}
\tag{48}
\end{equation}

where we have used the notation $\mathcal{F}\equiv e^{\frac{f}{2}}$ and $%
\Omega \equiv e^{\frac{h}{2}}$. The important thing here is that by setting $%
n=2$ and by performing the projections (41) on equation (33) we can obtain
(48) straightforward. Also from the action (46) one can obtain the field
equations for the case $b=0$, which correspond to the the well known higher
dimensional Schwarchild black-hole solution, namely

\begin{equation}
ds^{2}=-(1-\frac{k}{\varphi ^{D-1}(r)})(dx^{1})^{2}+\frac{d\varphi ^{2}}{(1-%
\frac{k}{\varphi ^{D-1}(r)})}+\varphi ^{2}(r)d^{D}\Omega .  \tag{49}
\end{equation}

Summarizing so far, by emphasizing that from the point of view of the
duality symmetry $a\rightarrow \frac{1}{a}$ and $b\rightarrow \frac{1}{b}$
the case $n=2$ seems to be exceptional, we have shown that the
Einstein-Hilbert action in $(n+D+d)-$dimensions is reduced to the action
(33) which for the case of $n=2$ contains the unified dynamics of both
cosmology and black-holes.

\bigskip

\section{\textbf{Non-Relativistic strings and the Einstein-Hilbert action.}}

\qquad Going further we would like to study the action (33) when $D=1$ and $%
d=1$. In this case, $\tilde{R}=0$ and $\hat{R}=0$ and therefore this action
becomes%
\begin{equation}
S=\int d^{2}x\sqrt{-\bar{g}}\{2\bar{g}^{AB}\partial _{A}a\partial _{B}b+(%
\bar{R}-2\Lambda )\}.  \tag{50}
\end{equation}

Let us write $a=\frac{1}{\sqrt{2}}\ln \lambda _{0}$ +$\frac{1}{\sqrt{2}}\ln
\lambda _{1}$ and $b=-\frac{1}{\sqrt{2}}\ln \lambda _{0}$ $+\frac{1}{\sqrt{2}%
}\ln \lambda _{1}$. One has $\partial _{A}a=\frac{(\lambda _{0})^{-1}}{\sqrt{%
2}}\partial _{A}\lambda _{0}+\frac{(\lambda _{1})^{-1}}{\sqrt{2}}\partial
_{A}\lambda _{1}$ and $\partial _{B}b=-\frac{(\lambda _{0})^{-1}}{\sqrt{2}}%
\partial _{B}\lambda _{0}+\frac{(\lambda _{1})^{-1}}{\sqrt{2}}\partial
_{B}\lambda _{1}$. Hence one finds that (50) can be rewritten as

\begin{equation}
S=\int d^{2}x\sqrt{-\bar{g}}\{-\bar{g}^{AB}\lambda ^{0}\lambda ^{0}\partial
_{A}\lambda _{0}\partial _{B}\lambda _{0}+\bar{g}^{AB}\lambda ^{1}\lambda
^{1}\partial _{A}\lambda _{1}\partial _{B}\lambda _{1}+(\bar{R}-2\Lambda )\}.
\tag{51}
\end{equation}

Here, $(\lambda _{0})^{-1}=\lambda ^{0}$ and $(\lambda _{1})^{-1}=\lambda
^{1}$. So, if once again one defines $x^{0}=\ln \lambda _{0}$ and $x^{1}=\ln
\lambda _{1}$ one sees that (51) can be rewritten as\bigskip 
\begin{equation}
S=\int d^{2}x\sqrt{-\bar{g}}\{-\bar{g}^{AB}\partial _{A}x^{0}\partial
_{B}x^{0}+\bar{g}^{AB}\partial _{A}x^{1}\partial _{B}x^{1}+(\bar{R}-2\Lambda
)\}.  \tag{52}
\end{equation}

It is remarkable that this action can be identified as the Polyakov action
for the string in $(1+1)$-dimensions. This result motivates us to generalize
(51) in the form

\begin{equation}
S=\int d^{2}x\sqrt{-\bar{g}}\{-\bar{g}^{AB}\lambda ^{0}\lambda ^{0}\partial
_{A}\lambda _{0}\partial _{B}\lambda _{0}+\bar{g}^{AB}\lambda ^{i}\lambda
^{j}\partial _{A}\lambda _{i}\partial _{B}\lambda _{j}+(\bar{R}-2\Lambda )\},
\tag{53}
\end{equation}%
with the indices $i,j,etc.$ running form $1$ to $p$. Consequently defining $%
x^{i}=\ln \lambda _{i}$ the action (53) becomes 
\begin{equation}
S=\int d^{2}x\sqrt{-\bar{g}}\{-\bar{g}^{AB}\partial _{A}x^{0}\partial
_{B}x^{0}+\bar{g}^{AB}\partial _{A}x^{i}\partial _{B}x^{j}\delta _{ij}+(\bar{%
R}-2\Lambda )\},  \tag{54}
\end{equation}%
which can be rewritten exactly as the Polyakov action in $(1+p)$-dimensions

\begin{equation}
S=\int d^{2}x\sqrt{-\bar{g}}\{ \bar{g}^{AB}\partial _{A}x^{\mu }\partial
_{B}x^{\nu }\eta _{\mu \nu }+(\bar{R}-2\Lambda )\}.  \tag{55}
\end{equation}%
Of course the term

\begin{equation}
\int d^{2}x\sqrt{-\bar{g}}\bar{R},  \tag{56}
\end{equation}
is classically a topologycal term, but it is necessary to consider at a
quantum level. Here we assume $\eta _{\mu \nu }=diag(-1,1,...,1)$ and $\mu
,\nu =0,1,2,...,p$.

\bigskip

\section{\textbf{The Gravi-dilaton effective action.}}

\qquad Thinking backwards one may ask, from which Einstein-Hilbert action
and line element can (53) be obtained? Let us consider now the gravi-dilaton
effective action

\begin{equation}
S=-\frac{1}{16\pi G_{d+1}}\int d^{d+1}y\sqrt{-g}e^{-\varphi }(R+\partial
\varphi \partial \varphi +2\Lambda ),  \tag{57}
\end{equation}%
where $G_{d+1}$ is the Newton constant in $d+1$ dimensions and $\varphi
=\varphi (y^{c})$ is the dilaton field. Moreover, $R$ is the Ricci scalar
obtained from the line element

\begin{equation}
ds^{2}=\bar{g}_{AB}(x^{C})dx^{A}dx^{B}+a_{k}(x^{C})a_{l}(x^{C})\eta
_{ij}^{kl}d\xi ^{i}d\xi ^{j},  \tag{58}
\end{equation}%
with $A,B,...=0,1,2,...,p+1$, and $i,j,...=p+2,p+3,...,d$, where the only
non-vanishing terms of $\eta _{ij}^{kl}$ are; $\eta _{ij}^{kl}=1$ when $%
k=l=i=j$.

\bigskip From (58) we have that

\begin{equation}
\begin{array}{c}
g_{AB}=\bar{g}_{AB}(x^{C}), \\ 
\\ 
g_{ij}=a_{k}(x^{C})a_{l}(x^{C})\eta _{ij}^{kl},%
\end{array}
\tag{59}
\end{equation}%
which leads us to the Ricci scalar $R=g^{\mu \nu }R_{\mu \nu }$ being

\begin{equation}
R=-2a^{i}\mathcal{D}^{A}\partial _{A}a_{i}-a^{i}a^{j}\partial
_{A}a_{i}\partial ^{A}a_{j}+a^{i}a^{j}\partial _{A}a_{k}\partial
^{A}a_{l}\eta _{ij}^{kl}+\bar{R}.  \tag{60}
\end{equation}

Then the action (57) becomes

\begin{equation}
\begin{array}{c}
S=-\frac{1}{16\pi G_{d+1}}\int d^{d+1}y\sqrt{-\bar{g}}\Pi a_{s}e^{-\varphi
}\{-2a^{i}\mathcal{D}^{A}\partial _{A}a_{i}-a^{i}a^{j}\partial
_{A}a_{i}\partial ^{A}a_{j} \\ 
\\ 
+a^{i}a^{j}\partial _{A}a_{k}\partial ^{A}a_{l}\eta _{ij}^{kl}+\partial
^{A}\varphi \partial _{A}\varphi +\bar{R}+2\Lambda \}.%
\end{array}
\tag{61}
\end{equation}

Also, consider the relation between $G_{d+1}$ and $G_{p+1}$,

\begin{equation}
\frac{1}{G_{p+1}}=\frac{V_{n}}{G_{d+1}},  \tag{62}
\end{equation}%
where $V_{n}$ is a volume constant in $n=d-p$ dimensions.

So we can rewrite the action (61) in the form

\begin{equation}
\begin{array}{c}
S=-\frac{1}{16\pi G_{p+1}}\int d^{p+1}y\sqrt{-\bar{g}}\mathcal{D}^{A}(-2\Pi
a_{s}e^{-\varphi }a^{i}\partial _{A}a_{i}) \\ 
\\ 
-\frac{1}{16\pi G_{p+1}}\int d^{p+1}y\sqrt{-\bar{g}}\Pi a_{s}e^{-\varphi
}\{a^{i}a^{j}\partial ^{A}a_{i}\partial _{A}a_{j}-2\partial ^{A}\varphi
a^{i}\partial _{A}a_{i} \\ 
\\ 
+\partial ^{A}\varphi \partial _{A}\varphi -a^{i}a^{j}\partial
^{A}a_{k}\partial _{A}a_{l}\eta _{ij}^{kl}+2\Lambda \} \\ 
\\ 
-\frac{1}{16\pi G_{p+1}}\int d^{p+1}y\sqrt{-\bar{g}}\Pi a_{s}e^{-\varphi }%
\bar{R},%
\end{array}
\tag{63}
\end{equation}%
where we have completed the total derivative of the first term, and since a
total derivative does not contribute to the dynamics of a classical system
we can drop it, and (63) becomes

\bigskip 
\begin{equation}
\begin{array}{c}
S=-\frac{1}{16\pi G_{p+1}}\int d^{p+1}y\sqrt{-\bar{g}}\Pi a_{s}e^{-\varphi
}\{a^{i}a^{j}\partial ^{A}a_{i}\partial _{A}a_{j}-2\partial ^{A}\varphi
a^{i}\partial _{A}a_{i} \\ 
\\ 
+\partial ^{A}\varphi \partial _{A}\varphi -a^{i}a^{j}\partial
^{A}a_{k}\partial _{A}a_{l}\eta _{ij}^{kl}+2\Lambda \} \\ 
\\ 
-\frac{1}{16\pi G_{p+1}}\int d^{p+1}y\sqrt{-\bar{g}}\Pi a_{s}e^{-\varphi }%
\bar{R}.%
\end{array}
\tag{64}
\end{equation}

\bigskip Now defining $x^{0}$ as

\begin{equation}
\Pi a_{s}e^{-\varphi }=e^{-x^{0}},  \tag{65}
\end{equation}%
we find

\begin{equation}
x^{0}=\varphi -\sum \ln a_{s}.  \tag{66}
\end{equation}

Using (66) we can rewrite the action (64) as

\bigskip 
\begin{equation}
\begin{array}{c}
S=\frac{1}{16\pi G_{p+1}}\int d^{p+1}y\sqrt{-\bar{g}}e^{-x^{0}}\{-\partial
^{A}x^{0}\partial _{A}x^{0}+a^{i}a^{j}\partial ^{A}a_{k}\partial
_{A}a_{l}\eta _{ij}^{kl}-2\Lambda \} \\ 
\\ 
-\frac{1}{16\pi G_{p+1}}\int d^{p+1}y\sqrt{-\bar{g}}e^{-x^{0}}\bar{R},%
\end{array}
\tag{67}
\end{equation}%
and defining the $p-brane$ coupling constant $\Omega _{p}$ in the form

\begin{equation}
\frac{e^{-x^{0}}}{16\pi G_{p+1}}=\frac{1}{2\Omega _{p}},  \tag{68}
\end{equation}%
the action (67) will take the form

\begin{equation}
\begin{array}{c}
S=\frac{1}{2}\int \frac{d^{p+1}y}{\Omega _{p}}\sqrt{-\bar{g}}\{-\partial
^{A}x^{0}\partial _{A}x^{0}+a^{i}a^{j}\partial ^{A}a_{k}\partial
_{A}a_{l}\eta _{ij}^{kl}-2\Lambda \} \\ 
\\ 
-\frac{1}{2}\int \frac{d^{p+1}y}{\Omega _{p}}\sqrt{-\bar{g}}\bar{R}.%
\end{array}
\tag{69}
\end{equation}

Also, by re-defining

\begin{equation}
x^{i}\equiv \ln a_{i},  \tag{70}
\end{equation}%
we have

\begin{equation}
\partial ^{A}x^{i}\partial _{A}x^{j}=a^{i}a^{j}\partial ^{A}a_{i}\partial
_{A}a_{j},  \tag{71}
\end{equation}%
so (69) becomes

\begin{equation}
\begin{array}{c}
S=\frac{1}{2}\int \frac{d^{p+1}y}{\Omega _{p}}\sqrt{-\bar{g}}\{-\partial
^{A}x^{0}\partial _{A}x^{0}+\partial ^{A}x^{i}\partial _{A}x^{j}\delta
_{ij}-2\Lambda \} \\ 
\\ 
-\frac{1}{2}\int \frac{d^{p+1}y}{\Omega _{p}}\sqrt{-\bar{g}}\bar{R},%
\end{array}
\tag{72}
\end{equation}%
which can be rewritten as

\begin{equation}
\begin{array}{c}
S=\frac{1}{2}\int \frac{d^{p+1}y}{\Omega _{p}}\sqrt{-\bar{g}}\{-\bar{g}%
^{AB}\partial _{A}x^{0}\partial _{B}x^{0}+\bar{g}^{AB}\partial
_{A}x^{i}\partial _{B}x^{j}\delta _{ij}-2\Lambda \} \\ 
\\ 
-\frac{1}{2}\int \frac{d^{p+1}y}{\Omega _{p}}\sqrt{-\bar{g}}\bar{R},%
\end{array}
\tag{73}
\end{equation}%
and finally we get (see Ref. [15])

\begin{equation}
S=\frac{1}{2}\int \frac{d^{p+1}y}{\Omega _{p}}\sqrt{-\bar{g}}\bar{g}%
^{AB}\partial _{A}x^{\hat{\mu}}\partial _{B}x^{\hat{\nu}}\eta _{\hat{\mu}%
\hat{\nu}}-\frac{1}{2}\int \frac{d^{p+1}y}{\Omega _{p}}\sqrt{-\bar{g}}(\bar{R%
}+2\Lambda ),  \tag{74}
\end{equation}%
where $\eta _{\hat{\mu}\hat{\nu}}=diag(-1,1,...,1)$ and $\hat{\mu},\hat{\nu}%
=0,1,2,...,n=d-p$.

\bigskip

As we can see so far, the first terms of the actions (56) and (74) have the
same form, with the exception of the constants and the number of
coordinates. On the action (56) we are considering a $2-$dimensional $g_{AB%
\text{ }}$metric while in (74) we are considering a more general case for $%
p+1$ dimensions.

\bigskip

\section{\textbf{Final remarks.}}

\bigskip

\qquad In this work we have been able to find a possible connection between
the FRW-cosmology, the Schwarzschild black-hole metric and string theory.
The main motivation for establishing these relations emerges from the fact
that by an appropriate coordinate transformation the static metric of the de
Sitter space can be mapped to a dynamical Friedmann-Robertson-Walker metric
with scale factor $a(T)=e^{\frac{T}{r_{0}}}$. It turns out that this static
and non-static relation between the de Sitter metric and the
FRW-cosmological model can be generalized also for $k=1$, $k=-1$ and their
corresponding scale factors $a(T)=\cosh \frac{T}{r_{0}}$ and $a(T)=\sinh 
\frac{T}{r_{0}}$, respectively [16] (See also Refs. [17] and [18]).

In this context we show that the FRW--cosmology and the black-hole
Schwarzschild metric can be considered as particular cases of a more general
constrained Lagrangian function in $(n+D+d)-$dimensions. These results
depend on the reductions applied to the $2-$dimensional metric,
corresponding to $n=2$. In both cases this means that one dimension of the $%
(n=2)-$subspace becomes a space-like coordinate.

An interesting aspect of our development is that the action (52) admits an
interpretation of the Polyakov action in 2 dimensions. Moreover, we also
show evidence that the procedure can be generalized to any dimension.

At this point it is important to highlight the main difference between
Kaluza-Klein theory and our approach. Kaluza-Klein method involves
compactification (dimensional reduction) of space-like coordinates while in
our formalism the time-like coordinate of the $2-$dimensional metric\ plays
an essential roll (see Ref.[19]).

For further research of the proposed unification of cosmology, black-holes
and string theory it would be interesting to consider quantum aspects. We
may expect that the quantum spectrum of string theory contains some
consequences on the quantum behavior of the dynamics of black-holes and
cosmology. It is tempting to consider the possibility that this formalism
could eventually lead to a different approach towards quantum gravity which
is still an open problem in physics (see Refs. [20] and [21] and references
therein).

Finally, it is important to mention that a correspondance between
black-holes and qubits in (4+4)-dimensions has already been shown in Ref.
[22], via the Cayley hyperdeterminant, which at the same time can be related
with oriented matroid theory (see Refs. [23]-[26] and references therein).
Furthermore, an 8-dimensional Ashtekar formalism has already been developed
as well (see Ref. [27]) which makes us wonder how interesting it would be,
as a further research, to relate this present work with all these recent
developments.

\bigskip

\bigskip

\pagebreak 

\bigskip

\bigskip \pagebreak

\end{document}